\documentclass[aps,amsmath,amssymb, twocolumn,superscriptaddress,prl
]{revtex4-1}
\usepackage{amssymb, amsmath}
\usepackage{bm}
\usepackage{graphicx}
\usepackage{color}
\usepackage{amsfonts}
\usepackage{latexsym}
\usepackage{subfigure}
\usepackage{ulem}
\usepackage{float}
\usepackage{tikz}
\usepackage{booktabs,array}
\usepackage{verbatim}
\usepackage{accents}
\usepackage{empheq}
\usepackage{mathtools}

\def\eq#1{{Eq.(\ref{#1})}}    \def\fig#1{{Fig.\ref{#1}}}

\usepackage{hyperref}

\begin{document}

\title{Contractility-driven cell motility against a viscoelastic resistance}
\date{\today}
\author{Tapas Singha}
\author{Pierre Sens}
\email{pierre.sens@curie.fr}
\affiliation{Institut Curie, PSL Research University, CNRS UMR 168; F-75005 Paris, France}

\begin{abstract}
We study a model of contraction-based cell motility inside a microchannel to investigate the regulation of cell polarization and motion by the mechanical resistance of the environment. A positive feedback between the asymmetry of the acto-myosin cortex density and cell motion gives rise to a spontaneous symmetry breaking beyond a threshold contractility that depends on the resistance of extracellular medium. In highly viscous environments, we predict bistability under moderate contractility, so that symmetry breaking needs to be activated. In a viscoelastic environment, we find periodic oscillations in cortex density and velocity polarization. At the boundary between viscous and viscoelastic environments, the cell may either cross into the viscoelastic medium, bounce back into the viscous medium, or become trapped at the boundary. The different scenarios defined different phase diagram that are confirmed by numerical simulations.  
\end{abstract}

\maketitle

Cell motility is a fundamental cellular mechanism crucial to many biological processes, from morphogenesis to cancer progression.
Cells move using a combination of protrusive forces generated by the polymerisation of polarised actin filaments and contractile forces, often concentrated in the cell cortex, a thin layer of actin and myosin motors underlying the cell membrane \cite{Bray, rafelski:2004}. Contractility-driven motility is particularly relevant for cells  in three-dimensional environments made by other cells or the extracellular matrix  \cite{Doyle09JCB, Doyle13rev,Poincloux11PNAS}. This phenotype is reproduced when cells are confined in microchannels \cite{liu:2015,ruprecht:2015,bergert:2015}, a popular setup to study how cell responds to geometrical constraints
\cite{Vargas14JoVE, *Rolli10POne, *Irimia09, *Irimia07LabChip, *Maiuri12}.

The cell cortex is a thin contractile layer of cytoskeleton of thickness $\simeq200 nm$ \cite{Clark13}, composed of multiple proteins: actin, myosin, cross-linkers, nucleators, etc. \cite{Biro2013}. Theoretical models based on the hydrodynamic theory of active gels \cite{marchetti:2013} have been developed to described its active properties
\cite{salbreux:2009, mayer:2010,bois:2011,joanny:2013,turlier:2014}, and applied to
contractility-based cell motility. In one class of model \cite{bergert:2015}, a predefined cell polarisation imposes a gradient of contractility which yields a cortical flow toward the more contractile region (the rear) of the cell. Friction between the flow and the environment generates traction forces that can move the cell. A second class of model \cite{hawkins:2011,callanjones13,Recho13PRL,ruprecht:2015} invokes an important feature of active contractile gels: the existence of a positive feedback loop between contractility gradients and cortical flows\cite{bois:2011}. As the flow advects  contractile elements toward more contractile regions of the cortex, the homogeneous state may be unstable, resulting in spontaneous cell polarisation and translocation without externally-imposed polarisation \cite{hawkins:2011,callanjones13,Recho13PRL,ruprecht:2015}.

The mechanical properties of the extra-cellular environment is known to affect the way cells can polarise and move \cite{yu:2011,*wolf:2011,*gu:2014}. 
Theoretical models predict that a cell with an imposed polarisation can move in a microchannel filled with a viscous fluid provided the friction between cell and channel wall exceeds a threshold proportional to viscous drag\cite{bergert:2015}. Here, we investigate how the cell's ability to exhibit spontaneous polarisation and motion depends on the mechanical resistance of the environment. 
We find that even in the limit of infinite friction, 
the motile steady state of broken symmetry only exists below a threshold of viscous drag coefficient. This could have strong physiological implications, as extracellular environments are quite often viscous. We further show that cortical dynamics qualitatively affects the ability of the cell to move in complex environments, such as viscoelastic environments, and mechanical boundaries between different extracellular media.

\begin{figure}[b]
\begin{center}
\includegraphics[width=8.5cm]{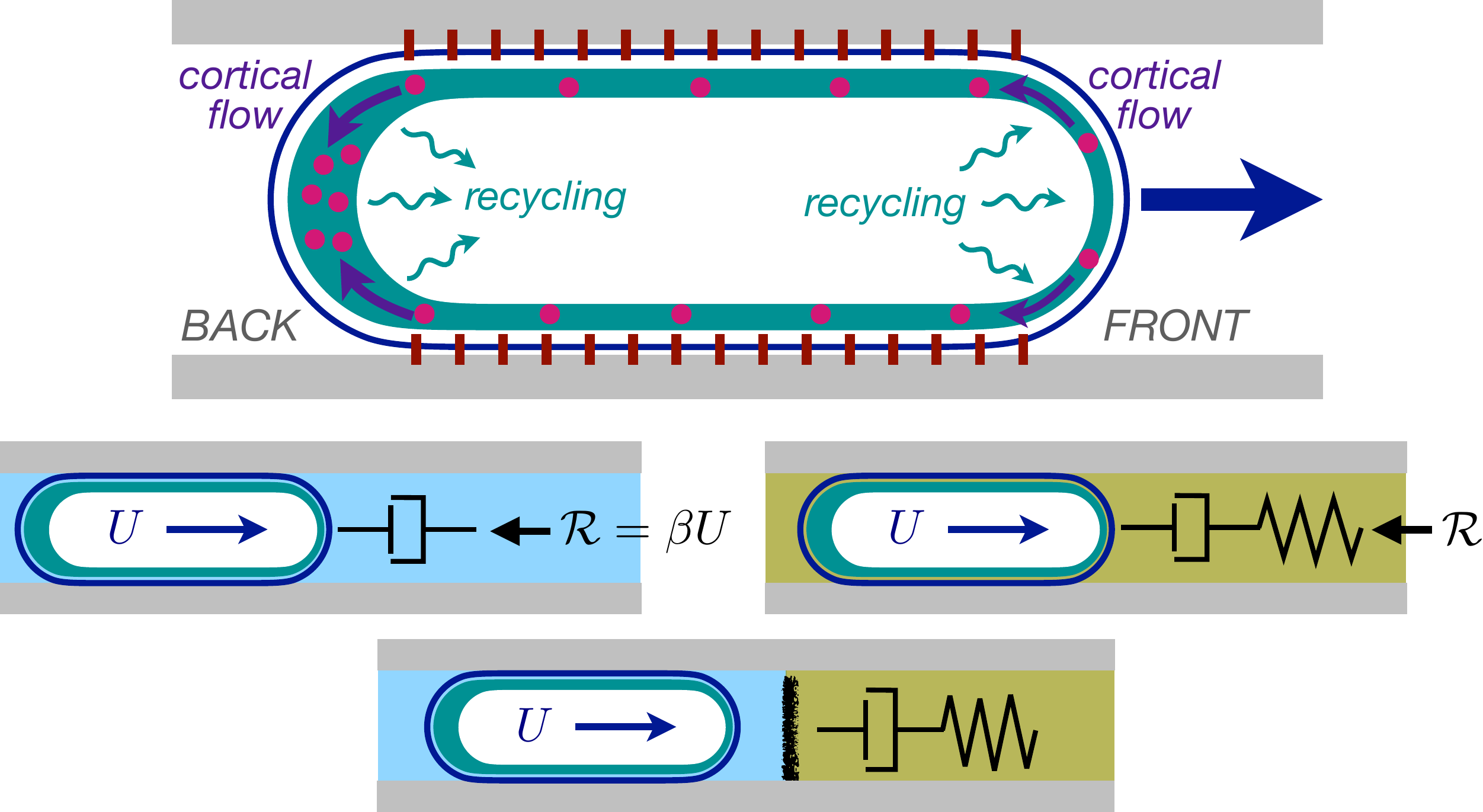}
\caption{\small {\sl No-slip cell motion in a microchannel.}
Top: Sketch of the model: motion is the consequence of spontaneous symmetry breaking leading to a cortex denser at the cell back and sparser at the cell front. This results in a difference of cortical tension that drive cortical flow. Cortex recycling (net depolymerisation at the back and net polymerisation at the front) transfers cortical matter forward, while the stress difference pushes the cytosol forward. Along the lateral surface in contact with the walls, the cortex is immobile with respect to the laboratory frame.  The  forward force $\Delta \gamma$ is balanced by a resistive force ${\cal R}$ which is either viscous, viscoelastic or the boundary between viscous and viscoelastic. } 
\label{fig1}
\end{center}
\end{figure}

{\bf Model.} We study a minimal model for the spontaneous polarisation and motion of a incompressible cell of fixed geometry confined in a circular channel (\fig{fig1}), wherein
gradients of contractility in the cytoskeleton cortex create cortical flow and generate friction forces with the channel walls. We neglect the friction between the cortical flow and the medium inside the channel at the cell edges, such that both the cortical tension $\gamma$ and the cortex curvature are constant at the two cell edges. Global force balance at the cell scale demands that $\Delta\gamma={\cal R}[U]$, where $\Delta\gamma=\gamma_b-\gamma_f$ is the tension difference between the cell edges (denoted back (b) and front (f)) and ${\cal R}$ is the resistive force (divided by the perimeter of the cell cross section) which depends on the cell velocity $U$. As detailed in the Supplementary Information (S.I.), it can stem from dissipation in the extracellular medium, in the cytoplasm, or from the detachment dynamics of adhesion molecules \cite{DecavePRL02, ward:1993, dimilla:1991}.

Following previous works \cite{bois:2011,Recho13PRL,ruprecht:2015}, we describe a simplified version of the model based on the one dimensional (1D) active fluid formalism, where the cortex is described by a single  density field, $\rho(x,t)$ that accounts for all protein populations, and a velocity field $v(x,t)$ describing cortical flow. 
The equation for density conservation reads 
 \begin{equation}
    \frac{\partial \rho(x,t)}{\partial t} +  \frac{\partial(\rho v)}{\partial x}  =  D \frac{\partial^2 \rho }{\partial x^2}  - k_d\left( \rho - \rho_0 \right)
\label{Eq_continuity}
\end{equation}
where $\rho_0$ is the reference (homeostatic) cortex density,  $D$ is a diffusion constant and  $k_d$ is the cortex recycling rate.

Various forms of  cortical stress tensor have been proposed in the literature \cite{salbreux:2009, mayer:2010,bois:2011,joanny:2013,turlier:2014,hawkins:2011,callanjones13,Recho13PRL,ruprecht:2015,bergert:2015}, and the specific formalism does not fundamentally alter the phenomenology given that the stress tensor $\gamma$ includes contractile and viscous terms, and higher order compressibility terms to ensure that teh density remains finite \cite{joanny:2013}. Here we adopt the following expression for a 1D compressible contractile fluid:
\begin{equation}
 \gamma (x,t) =   \zeta\, \left(\rho (x,t) - \frac{\rho^{2}(x,t)}{2 \rho^*} \right)   +   \eta  \frac{\partial}{\partial x}  v (x,t).
 \label{Eq_tension}
  \end{equation}
where $\zeta$ is the contractility, $\eta$ is the viscosity, and the non-linear saturating term is chosen to be quadratic with a characteristic density $\rho^*$.
The force balance of the cortex can be written as \cite{bois:2011,hawkins:2011,callanjones13,Recho13PRL,ruprecht:2015,bergert:2015} 
\begin{equation}
\frac{\partial\gamma (x,t)}{\partial x} =\Gamma\, v (x,t) 
\label{Eq_localFB}
\end{equation}
where $\Gamma$ is the friction coefficient. In the 1D model, the cell edges are of length $\ell$, fixed by the channel width, and the cell length in contact with the channel wall is denoted by $L$. In what follows, we neglect lateral friction at the cell edges ($\Gamma=0$ for $x\notin[\ell,L+\ell]$), 
an approximation valid for long cells (see S.I.),
and assume infinite friction with the walls ($\Gamma\rightarrow\infty$ for $\ell\le x\le L+\ell$), so that in the reference frame of the moving cell, the cortex velocity in contact with the walls is $v(x)=-U={\rm constant}$.

{\bf Viscous resistance: symmetry breaking and steady states}. Spontaneous cortical instability is driven by a mechanical feedback between contraction and flow, while recycling, diffusion and viscous dissipation stabilise an homogeneous cortex.
Linear stability analysis (LSA) (detailed in the S.I.) shows that, in the absence of friction, a uniform isolated slab of cortex of length $\ell$ is linearly unstable for $\bar\zeta > 1 + (\pi \ell_D)^2$ where $\bar\zeta = \zeta(1-\rho_0/\rho^*)/(\eta k_d)$  and $\ell^2_D=D/\ell^2 k_d$. Upon loss of stability, the cortical heterogeneities eventually saturate and coarsen due to non-linear terms  \cite{bois:2011}.

\begin{figure}[t]
\centering
\includegraphics[width=0.945\linewidth, height=0.265\textheight]{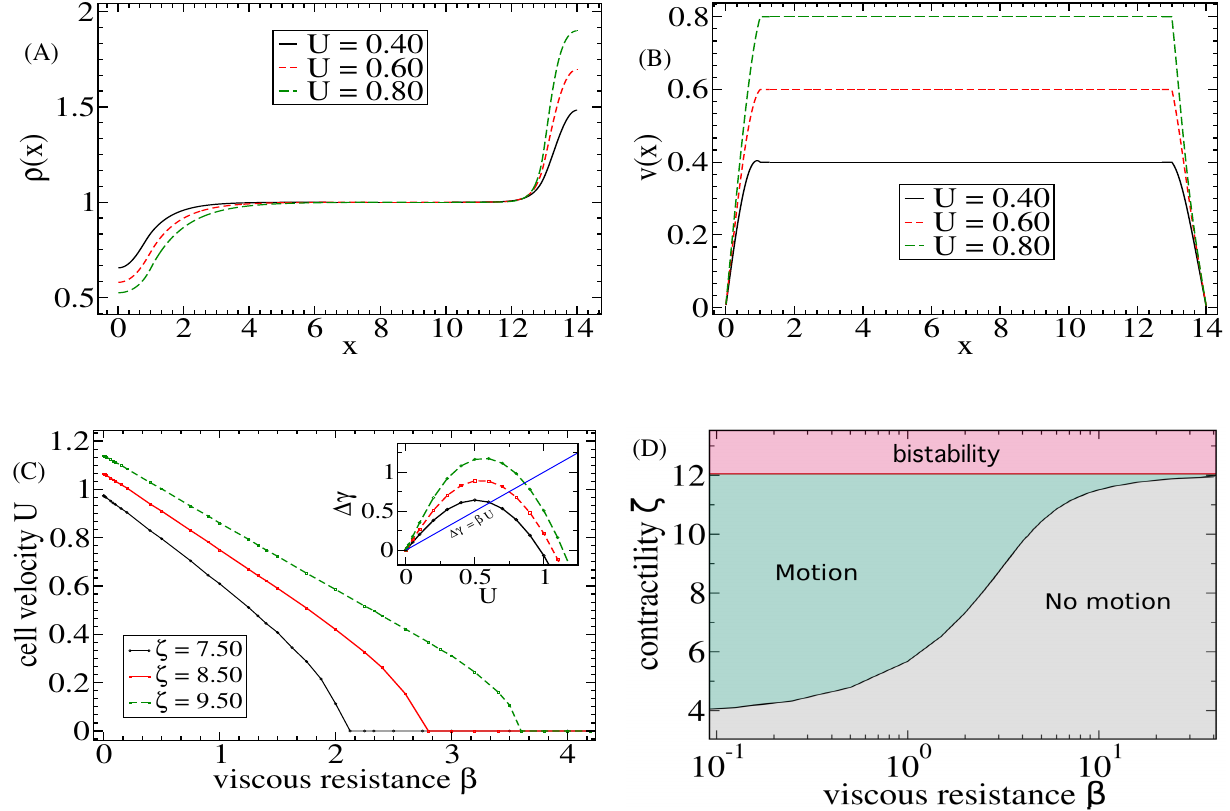}
\caption{{\sl Viscous environment}. {\bf A,B.} steady-state density (A) and cortex velocity (B) profiles. 
{\bf C.} Variation of the steady-state cell velocity  with the viscous resistance $\beta$ (inset is a graphical representation of the force balance \eq{Eq_GlobalFB}).  
{\bf D.} Viscous phase diagram showing the region where the cell spontaneously polarises and moves(teal). The threshold contractility (black line) - is drawn from the LSA. Above the red horizontal line is a region of bistability, showing the coexistence of a immobile state with a symmetric heterogeneous cortex, and a motile state with asymmetric cortex, which is stable even in the absence of motion (when $\beta\rightarrow\infty)$ (\eq{limit_visc}). 
With $\ell_D^2=0.25$ and $\rho^* = (3/2)\rho_0$.}  
\label{fig2}
\end{figure}

Under high friction with the channel walls, cell with broken symmetry moves in a caterpillar fashion, with an immobile lateral cortex (for $ \ell\leq x \leq L+\ell$) and a net rearward flux of cortical material balanced by polymerisation of the front ($x\leq\ell$, $\rho(x) <\rho_0$) and depolymerisation of the rear ($x\geq L+\ell$, $\rho(x)>\rho_0$) (\fig{fig1}). The steady-state solution of Eqs.(\ref{Eq_continuity},\ref{Eq_tension},\ref{Eq_localFB}) can be obtained numerically in the moving reference frame with appropriate boundary conditions: $v(0)=v(L+2\ell)=0$, $v(\ell)=v(L+\ell)=-U$, $\partial_x\rho(0)=\partial_x\rho(L+2\ell)=0$ and continuity of the flux at $x=\ell$ and $x=L+\ell$. Example of steady-state density and velocity profiles for different cell velocity are presented in \fig{fig2}(A) and (B), showing that the density gradient increase with cell velocity.

The cortical tension at each end of the cell is obtained from \eq{Eq_tension}. Enforcing global force balance then lead to a self-consistent determination of the steady-state cell velocity against a viscous environment using
\begin{equation}
 \Delta \gamma [U] = \mathcal R =  \beta\,U.
 \label{Eq_GlobalFB}
\end{equation}
where $\beta$ is an effective viscous drag and $U$ is the velocity of the cell. The variation of $\Delta\gamma$ with the cell velocity is shown as an inset in \fig{fig2}(C). Beyond a critical contractility, it shows a linear rise for small velocities, followed by a maximum and a decrease to negative values for high velocities. The self-consistent cell velocity solution of \eq{Eq_GlobalFB} is presented in \fig{fig2}C, showing that the motile steady-state only exist below a finite value of the viscous resistance $\beta$.

The critical value of contractility that gives rise to a motile state can be obtained from a LSA of the homogeneous immobile state, detailed in the S.I.
Briefly, Eqs.(\ref{Eq_continuity},\ref{Eq_tension},\ref{Eq_localFB}) are linearised considering a small perturbation about homogeneous density $\rho_0$ of a static cell ($U=0$) :
$\rho(x,t) = \rho_0 + \delta \rho(x,t)$ and $v(x,t) = \delta v(x,t)$.
In the cell edges we have:
$\frac{\partial \delta\rho}{\partial t} +\rho_0 \frac{\partial v}{\partial x} = D \frac{\partial^2 \delta\rho}{\partial x^2} - k_d\delta\rho$, and the local force balance equation is $\zeta(1-\rho_0/\rho^*) \frac{\partial \delta\rho}{\partial x} + \eta \frac{\partial^2 \delta v}{\partial x^2} = 0$
and in the middle part:
$\frac{\partial \delta\rho}{\partial t} = D \frac{\partial^2 \delta\rho}{\partial x^2} - k_d\,\delta\rho$ and $\delta v = -U$, with appropriate boundary conditions. 

Linear stability is investigated by expanding perturbations in density and velocity using eigenmodes that satisfy the boundary conditions. For very long cells (results for arbitrary length are discussed in the S.I.), these take the form: $ \delta\rho(x,t)=\sum_{n=1}^\infty \rho_ne^{\alpha_nt}[\cos{(q_n\,x)}+A_n]$ and $\delta v(x,t)=\sum_{n=1}^\infty v_n e^{\alpha_n t}[\sin{(q_n\,x)}+B_n x]$ at the cell edges and $ \delta \rho(x,t)=\sum_{n=1}^\infty b_n\,e^{\alpha_nt}e^{-p_n (x-l)}$ and $\delta v(x,t)=-U$ for the region in contact with the walls, where $n$ is the mode index and $\alpha_n$ is the growth rate which is to be determined. 

The linear stability analysis predicts the existence of a contractility threshold $\overline{\zeta^*}$ beyond which there exists at least one unstable mode (see S.I.). The relation of the threshold contractility with the viscous resistance is:
\begin{equation}
\tan \left(\frac{\sqrt{\overline{\zeta^*}-1}}{\ell_D}\right) = \frac{\sqrt{\overline{\zeta^*}-1}}{\overline{\zeta^*}-1 - \frac{\overline{\zeta^*}\, \frac{\bar\beta}{2} \ell_D}{\overline{\zeta^*}-(1+\frac{\bar\beta}{2})}}.
\label{Eq_thresholdZeta}
\end{equation}
with $\bar\beta=\beta\ell/\eta $. The numerical solution of this equation is shown as a function of the viscous resistance $\beta$ in the motility phase diagram of \fig{fig2}(D) (black curve). It separates the parameter region where the homogeneous state is stable and no cell motion is expected ($\zeta<\zeta^*$) and the region where the cell spontaneously polarises and moves ($\zeta>\zeta^*$). The asymptotic solutions of \eq{Eq_thresholdZeta} for low and high resistance are:
\begin{equation}
\overline{\zeta^*}\xrightarrow[\beta\rightarrow0]{}1+\left(\frac{\pi}{2}\ell_D\right)^2\qquad
\overline{\zeta^*}\xrightarrow[\beta\rightarrow\infty]{}1+\left(a\pi\ell_D\right)^2
\label{limit_visc}
\end{equation}
with $1<a<3/2$. The first limit corresponds to the threshold obtained for an isolated slab of twice the length of one cell edge. The second limit (red line in \fig{fig2}(D)) is in fact relevant for any $\beta$ value. Above this threshold, cortex heterogeneities can develop and reach a steady state independently at the two cell edges. The two profiles may be symmetrical with respect to the cell center, in which case the cell does not move, or non-symmetrical, which allows for cell motion with a velocity dependent upon $\beta$. The steady-state is thus bistable and the cell velocity in the motile state decreases continuously to zero as $\beta$ increases to infinity. This highly contractile behaviour is discussed in the S.I.

{\bf Viscoelastic resistance.} While physiological extracellular media often appear rather viscous at long times, their short-time response is generally elastic. One may expect  the interplay between the intrinsic dynamics of the cortex and the relaxation dynamics of the medium to give rise to interesting features. We investigate this by considering that  the channel is filled by a simple Maxwell viscoelastic (VE) material of stiffess $k$ and relaxation time $\tau$, for which the global force balance reads:
 \begin{equation}
   \label{eq:viscoel_force_bal}
\frac{d \Delta \gamma}{d t}  + \frac{\Delta \gamma }{\tau} = k \,U.
 \end{equation}
The LSA may be performed using the same eigenmodes as before, but the presence of a time derivative in \eq{eq:viscoel_force_bal} results in qualitative differences, as the perturbation growth rate ($\alpha_n$) may be complex. The lowest mode ($n=1$) is always the first to become unstable, and its growth rate satisfies $\alpha_1=\bar{\zeta}-1-q^2$, where $q=\ell_Dq_1$ satisfies 
\begin{equation}
\tan\left(\frac{q}{\ell_D}\right) = \frac{R(q)\,q\, \sqrt{\overline{\zeta}-q^2}}{R(q)\, q^2 -\frac{\bar k\,\bar\tau \overline{\zeta} \ell_D }{2 q^2} \sqrt{\overline{\zeta}-q^2}}
\label{Eq_CondiVE}
\end{equation}
where $R(q)=(1+\frac{\bar k\,\bar\tau}{2}-\bar\tau)+ \bar\tau\, \overline{\zeta}\, \left(1-\frac{k}{2 q^2}\right)-\bar{\tau} q^2$, with $\bar\tau=k_d\tau$ and $\bar k=k\ell/(\eta k_d)$. Solving both the real and imaginary parts of \eq{Eq_CondiVE}, with $q_1$ a complex number and requiring the condition for marginal stability $Re[\alpha_1]=0$ fully determines the critical contractility $\zeta^{**}$ for instability in a VE environment.

\begin{figure}[b]
\centering
\includegraphics[width=0.94\linewidth, height=0.30\textheight]{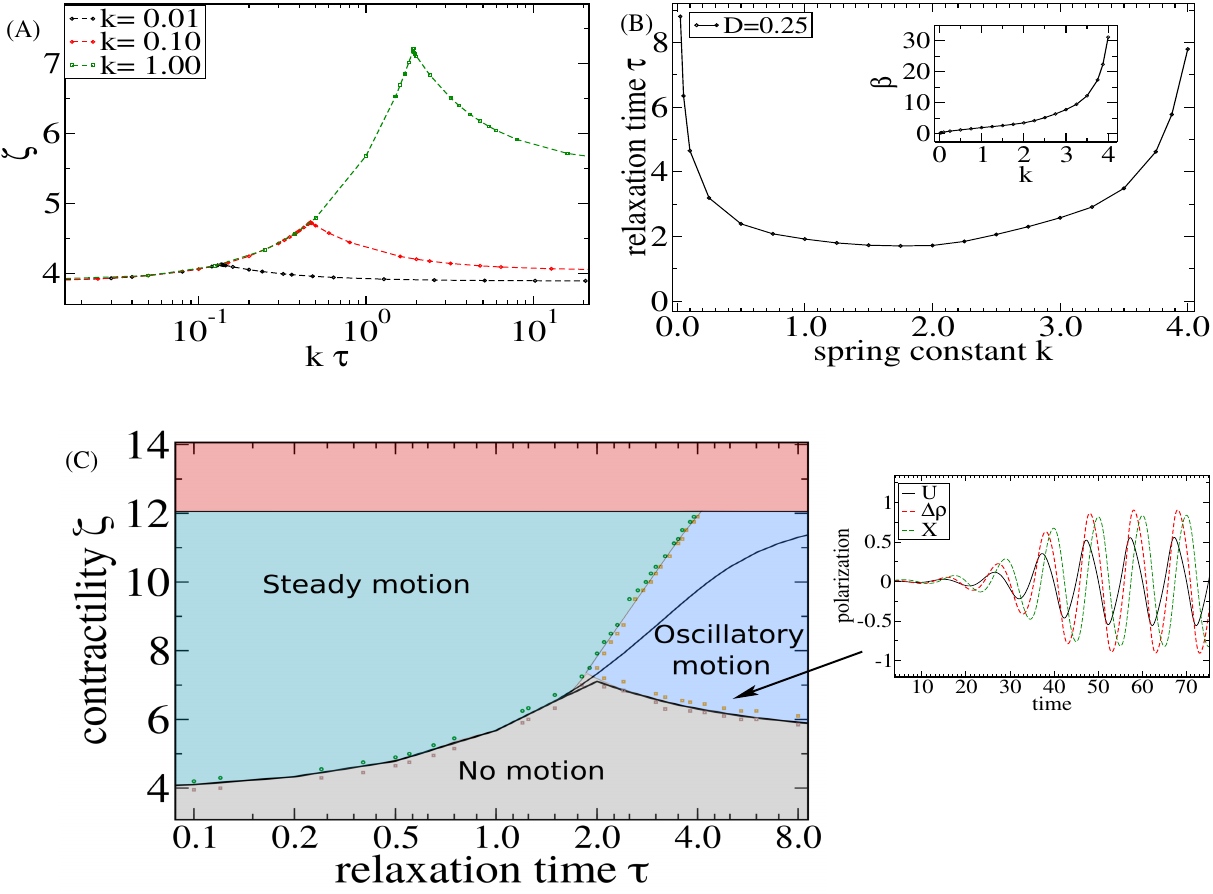}
\caption{{\sl Viscoelastic resistance}. A. Critical VE contractility threshold solution of \eq{Eq_CondiVE} for different  values of $k$. The kink in the curves indicates the VE relaxation time beyond which eigenmodes are complex, leading to oscillatory motion. B. Value of the relaxation time at which the short-time elastic response influences motion (kink) as a function of $k$. C. Phase diagram of cell motion. The critical contractility in viscous environment is shown as a dashed line. Oscillatory motion is observed for long VE relaxation time. In this regime, the transition to oscillatory motion is captured by the linear stability analysis, but the boundary between steady and oscillatory motion is inherently non-linear and requires numerical resolution of the full set of spatio-temporal equations (Eqs.(\ref{Eq_continuity},\ref{Eq_tension},\ref{eq:viscoel_force_bal}) - dots indicate numerical results). Inset: examples of non-linear periodic variation of the cell velocity, position and edge density difference in the oscillatory regime. Parameters are $\rho^* = 3/2$, $\ell_D^2=0.25$ and $\overline{k}=1$ for the bottom panel.
} 
\label{fig3}
\end{figure}

The numerical solutions of \eq{Eq_CondiVE} is shown in the viscoelastic phase diagram of \fig{fig3}. For fast VE relaxation ($k_d\tau$ smaller than a threshold shown in \fig{fig3}B), $\alpha_1$ is real and the VE threshold contractility is identical to the viscous threshold with a viscous resistance $\beta=k\tau$. Physically, this is because the viso-elastic relaxation rate is faster than the rate at which concentration heterogeneities develop, so that the short-time elastic response has no influence. Mathematically, \eq{Eq_thresholdZeta} is strictly equivalent to \eq{Eq_CondiVE} when its solution is real. Motile cells then display steady motions equivalent to the viscous case. On the other hand, for slow VE relaxation, the short time elastic force resisting cell motion  drives the reversal of polarity. The unstable growth rate is then complex and leads in permanent oscillations of both cortical asymmetry and cell position, resulting in a non-linear oscillatory motion. 

\begin{figure}[t]
\centering
\includegraphics[width=1\linewidth]{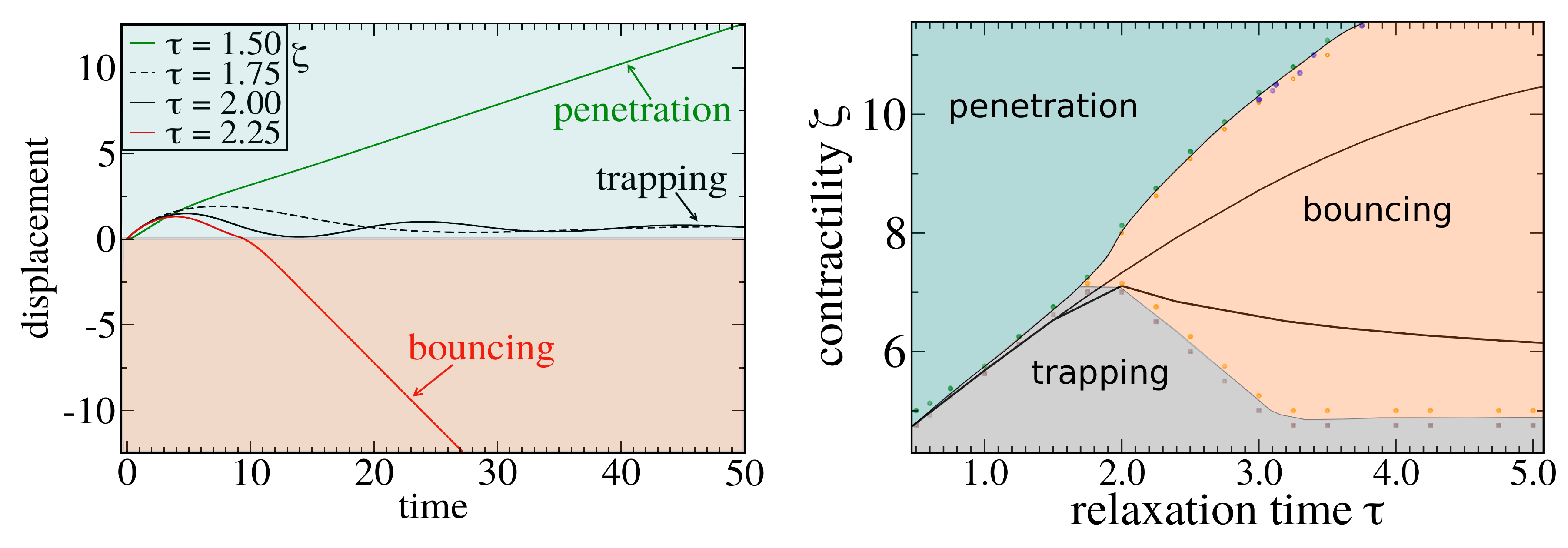}%
\caption{\small {\sl Cell penetrating a viscoelastic medium.} (left) Example for cell trajectories at the boundary between a viscous fluid (pink: $\beta_0=0.5$) and a visco-elastic fluid (green: $k=1$), showing the three different behaviours observed for different VE relaxation times: penetration, trapping and bouncing. (right) Phase diagram for the three types of behaviour varying the cell contractility and the VE relaxation time. 
} 
\label{fig4}
\end{figure}

{\bf Viscoelastic boundary.} 
Cells migrating through physiological tissues encounter  mechanically heterogeneous environments. An idealised version of this can be studied by analysing the behaviour of a motile cell encountering a viscoelastic boundary. This situation involves the  non-linear response of the cell and must be studied numerically. Examples of the cell trajectory at the boundary are shown in \fig{fig4}(A) and the phase diagram for the long-time cell behaviour in the ($\tau,\zeta$) space is shown in \fig{fig4}(B). For fast VE relaxation, the behaviour corresponds to that at the boundary between two viscous media with different viscous resistance and can be understood from the viscous phase diagram (\fig{fig2}). The cell is able to penetrate if its contractility is above the viscous threshold of the  medium it encounters, or gets trapped at the boundary otherwise. In the latter case, the dynamics evolution toward the symmetric state is over-damped and no reversal of symmetry and escape from the boundary is possible without noise. For slower VE relaxation time, the short-time elastic nature of the boundary leads to the existence of a bouncing state. The cell movement and polarisation may be reversed if the elastic restoring force is slow to relax, in which case the cell changes direction and bounces off the boundary back into the viscous medium. This bouncing state becomes prominent as $\tau$ increases, making it unlikely for a cell to penetrate a viscoelastic medium with a relaxation time much slower than $1/k_d$. 

{\bf Discussion.} 
The mechanical properties  of the environment plays a crucial role in regulating the ability of cells to invade new tissues \cite{yu:2011,*wolf:2011,*gu:2014}. While cellular mechano-sensing involves complex molecular pathways \cite{disher:2005}, pure mechanical response does have a direct effect on the cell polarisation. We use a simple 1D model of a very long cell in a confined geometry to study how the mechanical resistance of the environment influences contractility-driven cell motility. A positive feedback between local density variations and flows in the cellular cortex leads to  spontaneous symmetry breaking and cell polarisation.
Friction between the cortical flow and the confining walls generates a propulsion force that allows cell motion. We restrict ourselves to the limit of strong friction, and use linear stability analysis to predict a variety of cell behaviours depending on
the mechanical resistance of the external medium, including spontaneous motion, bistability where steady motion needs to be activated, and oscillatory behaviour. 
Numerical simulations support these predictions, and allow us to compute the non-linear cell behaviour at the boundary between two elastically distinct media. These result show that a cell may not easily penetrate a viscoelastic medium with a relaxation time slower than  the typical turnover time of the cellular cortex.   

Our results are expressed in terms of dimensionless parameters: primarily the cell contractility $\bar\zeta\propto\zeta\rho_0/(\eta k_d)$, the normalised friction  $\bar\beta\equiv \beta l/\eta$ and relaxation time $\bar\tau \equiv k_d\tau$ of the extracellular medium. Using typical values of the parameters (cortical tension: $\zeta\rho_0=10^{-4}$ N/m, cortex viscosity: $\eta=10^{-3}$ Pa.m.s and recycling rate: $k_d\simeq 0.01-0.1$ s$^{-1}$ \cite{salbreux:2012,Fritzsche14,saha:2016}) one obtains a value of the dimensionless contractility $\overline{\zeta}\simeq 1-10$, well within the range of spontaneous symmetry breaking.
The effective viscoelastic parameters $\beta$ and $k$ are related to the viscosity and Young's modulus of the extracellular medium  in complex ways, that depend on the boundary conditions at the channel walls and may involve the channel radius and length (see S.I.) \cite{bergert:2015}. With $\beta \sim$ viscosity $\simeq 10^3-10^4$ Pa.s \cite{stirbat:2013} and Young's modulus for soft tissues $\simeq 0.1-1$ kPa \cite{gu:2014}, one finds $\bar\beta\simeq 1-10$  and $\bar\tau\simeq 0.1-10$.
Thus the expected parameter values are within the regions of interest of the phase diagrams (Figs.\ref{fig2},\ref{fig3},\ref{fig4}) establishing the physiological relevance of our results.

\begin{acknowledgements}
We acknowledge funding from: INSERM ITMO cancer 20CR110-00
\end{acknowledgements}

\bibliographystyle{unsrt}
%

\bibliographystyle{apsrev4-1}
\end{document}